\documentclass[showpacs,aps,twocolumn]{revtex4}

\usepackage{bm}
\usepackage{amsmath}
\usepackage{graphicx}
\usepackage{subfigure}
\usepackage[usenames,dvipsnames]{color}
\definecolor{darkblue}{RGB}{0,0,196}
\usepackage[colorlinks=true,linkcolor=darkblue,citecolor=darkblue,urlcolor=darkblue]{hyperref}

\usepackage{setspace}
\usepackage{footmisc}
\usepackage[makeroom]{cancel}
\usepackage{comment}

\usepackage{color,soul}

\def\be{\begin{equation}}
\def\ee{\end{equation}}
\def\ba{\begin{eqnarray}}
\def\ea{\end{eqnarray}}

\usepackage{graphicx}
\usepackage{amsmath,bbm}
\usepackage{amssymb,bm}

\begin{document}

\title{Energy and Centrality Dependent Study of Deconfinement Phase Transition in a Color String Percolation Approach at RHIC Energies}
\author{Pragati~Sahoo}
\author{Sudipan De}
\author{Swatantra~Kumar~Tiwari}
\author{Raghunath~Sahoo\footnote{Corresponding Author Email: $Raghunath.Sahoo@cern.ch$}}
\affiliation{Discipline of Physics, School of Basic Sciences, Indian Institute of Technology Indore, Simrol, Indore- 453552, INDIA}

\begin{abstract}
\noindent

We take the experimental data for transverse momentum spectra of identified charged hadrons in different centrality classes for nucleus-nucleus (A+A) collisions at various Relativistic Heavy-Ion Collider (RHIC) energies measured by the STAR collaboration. We analyse these data in the framework of color string percolation model (CSPM) in order to extract various percolation parameters at different centralities at RHIC energies to study the effect of collision geometry and collision energy. We use these parameters to study the centrality dependent behaviour of initial temperature of the percolation cluster, energy density, average transverse momentum, shear viscosity to entropy density ratio ($\eta/s$) and trace anomaly for different energies at RHIC from $\sqrt{s_{\rm NN}}$ = 19.6 to 200 GeV. These observables are found to strongly depend on centrality at various collision energies. The critical percolation density, which is related to the deconfinement phase transition is achieved in the most central nucleus-nucleus collisions, while it fails in the peripheral collisions. A universal scaling is observed in color suppression factor and initial temperatures when studied as a function of charged particle pseudorapidity distribution scaled by nuclear overlap area at RHIC energies. The minimum of $\eta/s$ is observed in the most central collisions at $\sqrt{s_{\rm NN}}$ = 130 and 200 GeV.

\end{abstract}

\date{\today}
\maketitle 
\section{Introduction}
\label{intro}

The color string percolation model (CSPM) describes the initial collision of two heavy ions in terms of color strings stretched between the projectile and target. Color strings may be viewed as small discs in the transverse space filled with the color field created by colliding partons~\cite{prc65, epjc16}. Particles are produced by the Schwinger mechanism, emitting $\it q\bar q$ pairs in this field~\cite{taopro}.  With the growing energy and size of the colliding nuclei the number of strings grow and start to overlap to form clusters in the transverse plane. This procedure is very much like discs in the 2-dimensional percolation theory, a non-thermal second order phase transition~\cite{epjc71}. At a certain critical string density a macroscopic cluster appears, which defines the percolation phase transition~\cite{Phyreport}. In CSPM the Schwinger barrier penetration mechanism for particle production and the fluctuations in the associated string tension due to the strong string interactions make it possible to define temperature. The critical density of percolation is related to the effective critical temperature and thus percolation may provide the information of the deconfinement in the heavy-ion collisions~\cite{PLB642}. The CSPM has been successfully used to describe the initial stages in the soft region of the high energy heavy-ion collisions~\cite{Phyreport}. 

The aim of this paper is to study the centrality dependence of percolation parameters at various RHIC energies ranging from $\sqrt{s_{\rm NN}}$ = 19.6 to 200 GeV using CSPM. We use these parameters to calculate the centrality dependence of thermodynamical observables such as initial temperature of the percolation cluster, energy density, average transverse momentum, shear viscosity to entropy density ratio, and trace anomaly at RHIC energies. The paper runs as follows. First, we present the formulation and methodology in section~\ref{Form}. In section~\ref{RD}, we present the results and discussions. Finally, the findings of this work are summarized in section~\ref{summary}.

\section{Formulation And Methodology}
\label{Form}

In CSPM, the hadron multiplicity ($\mu_{\rm n}$) reduces as the interactions of strings increase and the mean transverse momentum squared, $\langle p_{T}^2 \rangle_{\rm n}$ of these hadrons increases, to conserve the total transverse momentum. The hadron multiplicity and $\langle p_{T}^2 \rangle_{\rm n}$ are directly related to the field strength of the color sources and thus to the generating color. For a cluster of $n$ individual strings, we have~\cite{epjc71} 

\begin{eqnarray}
n = \frac{\mu_{\rm n}}{\mu_0}\frac{\langle p_{T}^2 \rangle_{\rm n}}{\langle p_{T}^2 \rangle _1 },
\end{eqnarray}

where $\mu_0$ and $\langle p_{T}^2 \rangle _1$ are the multiplicity of hadrons and the mean transverse momentum squared of a single string, respectively.

The multiplicity, $\mu_{\rm n}$ and the mean transverse momentum squared $\langle p_{T}^2 \rangle_{\rm n}$ of the particles produced by a cluster of $n$ strings is defined as~\cite{Phyreport}:

\begin{eqnarray}
\mu_{\rm n} =\sqrt{\frac{nS_{\rm n}}{S_{1}}}\mu_{0},
\label{eq1}
\end{eqnarray}

\begin{eqnarray}
\langle p_{T}^2 \rangle_{\rm n} = \sqrt{\frac{nS_{1}}{S_{\rm n}}}\langle p_{T}^2 \rangle_{1},
\end{eqnarray}
where $S_{1}$ is the transverse area of a single string. $S_{\rm n}$ is the transverse area covered by $n$ strings. The percolation density parameter ($\xi$) can be written as:
 
 \begin{eqnarray}
\xi = \frac{N_{S}S_1}{S_n} ,
\label{eq2}
\end{eqnarray}
where $N_S$ is the number of strings formed in the nuclear collisions.

$\xi$ is evaluated by using the parameterisation of pp collisions at $\sqrt{s}$ = 200 GeV as it was performed in our earlier work~\cite{Sahoo:2017umy}. The $p_{T}$ spectrum of charged particles can be described by a power law:

\begin{eqnarray}
\frac{d^{2}N_{\rm ch}}{dp_{T}^{2}} = \frac{a}{(p_{0}+{p_{T}})^{\alpha}},
\label{eq3}
\end{eqnarray}

where $a$ is the normalisation factor and  $p_{0}$, $\alpha$ are fitting parameters given as, $p_{0}$ = 1.982 and $\alpha$ = 12.877~\cite{Braun:2015eoa}. Due to the low string overlap probability, the fit parameters obtained from pp collisions are then used for Au+Au collisions in order to evaluate the interactions of the strings.

In order to include the interactions of strings in nucleus-nucleus collisions, the parameter $p_{0}$ is modified as~\cite{Phyreport},

\begin{eqnarray}
 p_0\rightarrow p_0\left(\frac{\langle  nS_1/S_n\rangle_{\rm Au+Au}}{\langle nS_1/S_n\rangle_{pp}}\right)^{1/4}.
 \label{eq4}
\end{eqnarray} 

Using thermodynamic limit, {\it i.e.} $n$ and $S_n \rightarrow \infty$ and keeping $\xi$ fixed, we get

\begin{eqnarray}
 \langle \frac{nS_{1}}{S_{\rm n}} \rangle = \frac{1}{F^{2}(\xi)},
 \label{eq5}
\end{eqnarray}

where $F(\xi)$ is the color suppression factor which reduces the hadron multiplicity from $n\mu_0$ to the interacting string value, $\mu$ as, 

\begin{eqnarray}
\mu = F(\xi)n\mu_0,
\end{eqnarray}

where,

\begin{eqnarray}
 F(\xi) = \sqrt \frac{1-e^{-\xi}}{\xi}.
 \label{eq6}
\end{eqnarray}

Using eqs.~\ref{eq3} and~\ref{eq5}, we calculate for A+A collisions as,
\begin{eqnarray}
\frac{d^{2}N_{\rm ch}}{dp_{T}^{2}} = \frac{a}{(p_{0} \sqrt {{F(\xi)_{pp}/F(\xi)}_{\rm Au+Au}^{cent}}+{p_{T}})^{\alpha}}.
\label{eq7}
\end{eqnarray}

Here $F(\xi)_{Au+Au}^{cent}$ is the centrality dependent color suppression factor and $F(\xi)_{pp} \sim 1$ at low energies due to the low overlap probability.

The initial temperature of the percolation cluster, $T(\xi)$ can be represented in terms of F($\xi$) as~\cite{Phyreport}: 
\begin{eqnarray}
T(\xi) = \sqrt \frac{\langle p_{T}^{2} \rangle_{1}}{2F(\xi)}.
\label{eq8}
 \end{eqnarray}	
 
The mean transverse momentum squared of a single string, $\langle p_{T}^{2} \rangle_{1}$ is calculated using eq.~\ref{eq8} at critical temperature, $T_{c}$ = 167.7 $\pm$ 2.76 MeV \cite{Becattini:2010sk} and $\xi_c$ = 1.2. We get $\sqrt{\langle p_{T}^{2} \rangle_{1}}$ = 207.2$\pm$3.3 MeV~\cite{Phyreport}, which is $\simeq$ 200 MeV, obtained in the previous calculation using percolation model~\cite{PLB642}. 
The initial temperature of the percolation cluster is in reasonable agreement with the initial temperature estimated from the direct photon measurement by the PHENIX collaboration at $\sqrt{s_{\rm NN}}$ = 200 GeV~\cite{Adare:2008ab}.

\section{Results and Discussions}
\label{RD}
In the present work, we have extracted $\xi$ and $F(\xi)$ at mid-rapidity ($|y|<$ 0.5) for different centrality classes using the transverse momentum spectra of charged particles produced in Au+Au collisions at RHIC energies from $\sqrt{s_{\rm NN}}$ = 19.6 to 200 GeV~\cite{star1,star2, Adler:2002xw,Abelev:2007ra}. 

\begin{figure}
\includegraphics[height=20em]{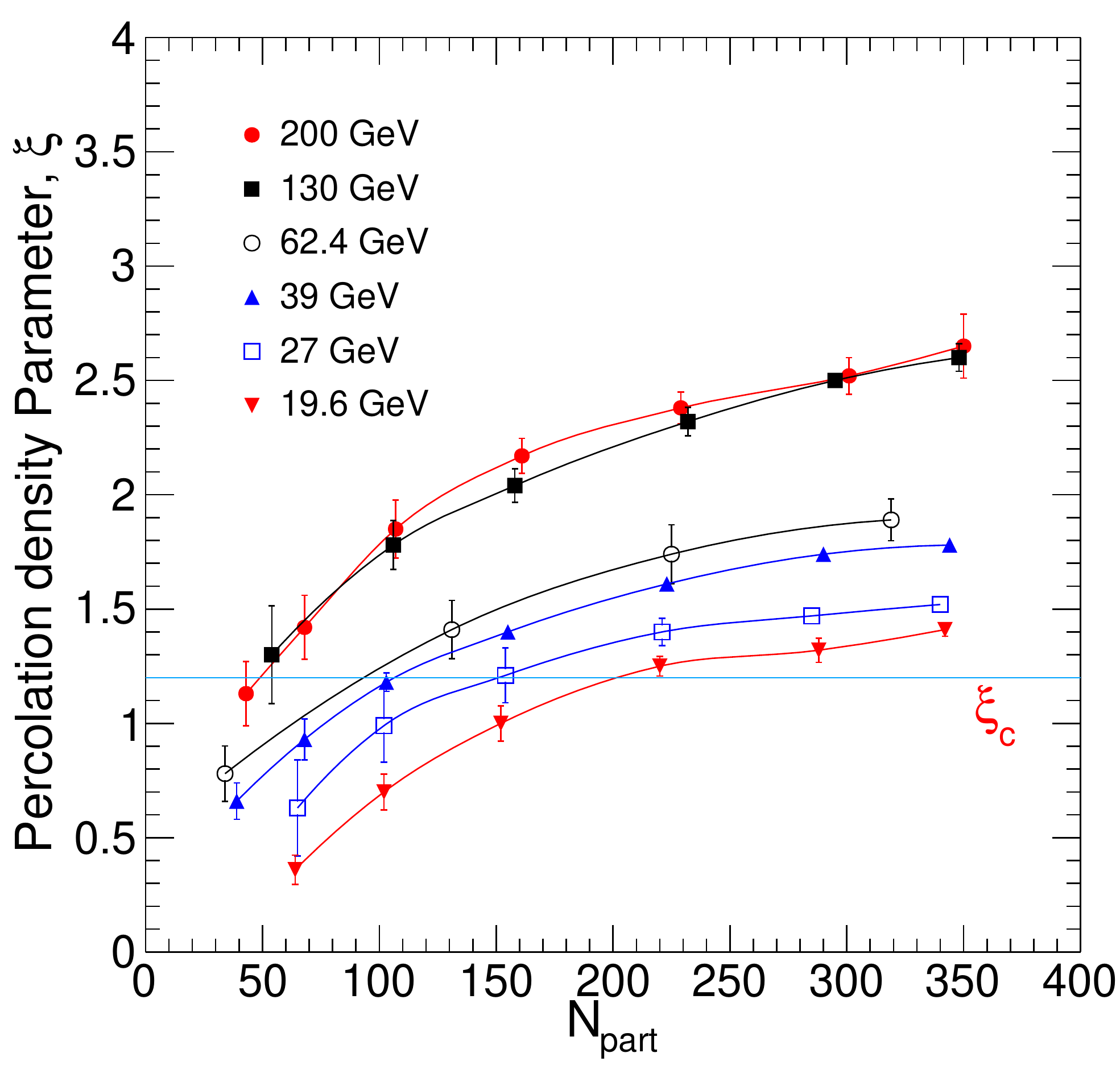}
\caption[]{(Color online) Percolation density  parameter $\xi$,  as a function of number of participants ($N_{part}$) for Au+Au collisions at RHIC energies from 19.6 - 200 GeV. The symbols correspond to different centre-of-mass energies and the horizontal line is the critical percolation density.}
\label{xi_Np}
\end{figure}

Figure~\ref{xi_Np} shows the extracted percolation density parameter, $\xi$ as a function of number of participants ($N_{part}$) in Au + Au collisions for $\sqrt{s_{\rm NN}}$ = 19.6 to 200 GeV. The proton to pion ratio is affected by baryon stopping at lower center-of-mass energies. We do not consider the results of percolation density parameter for $\sqrt{s_{\rm NN}}$ = 7.7 and 11.5 GeV, as for these energies the proton to pion ratios are much larger compared to higher collision energies.

The $\xi$ values are obtained by fitting the experimentally measured transverse momentum spectra of charged particles. $N_{part}$ values for different centrality classes are obtained by using Glauber model calculation~\cite{Loizides:2016djv}. The uncertainties in $\xi$ are estimated by varying the fitting conditions. The associated systematic uncertainties in percolation density parameters are $\sim$ 3\%. We have propagated the uncertainties for other observables accordingly.

 It is observed that $\xi$ values are higher for higher centre-of-mass energies and increase with $N_{part}$ for a particular energy. This indicates that the string overlap is greater for central collisions. The horizontal line corresponds to the critical percolation density parameter, $\xi_{c} \sim$ 1.2. The critical density for the onset of continuum percolation has been determined in numerical studies for a variety of different systems. In 2-dimensions the threshold for percolation is $\xi_{c} \sim$ 1.2~\cite{Satz:1998kg,Satz:2000bn,Isichenko:1992zz}.
 It is observed that $\xi_{c} >$ 1.2 for $N_{part} >$ 50 at $\sqrt{s_{\rm NN}}$ = 130 and 200 GeV. Whereas, for the lower collision energies starting from $\sqrt{s_{\rm NN}}$ =  19.6 - 62.4 GeV, $\xi_{c} >$ 1.2 for central collisions. This suggests that the percolation phase transition is expected  for $N_{part} >$ 50 at $\sqrt{s_{\rm NN}}$ = 130 and 200 GeV and for central collisions at lower energies. This is in conformity with the higher effective energy available for particle production in central collisions~\cite{Sarkisyan,Mishra:2014dta,Sarkisyan-Grinbaum:2018yld,Sarkisyan:2015gca}.  

\begin{figure}
\includegraphics[height=20em]{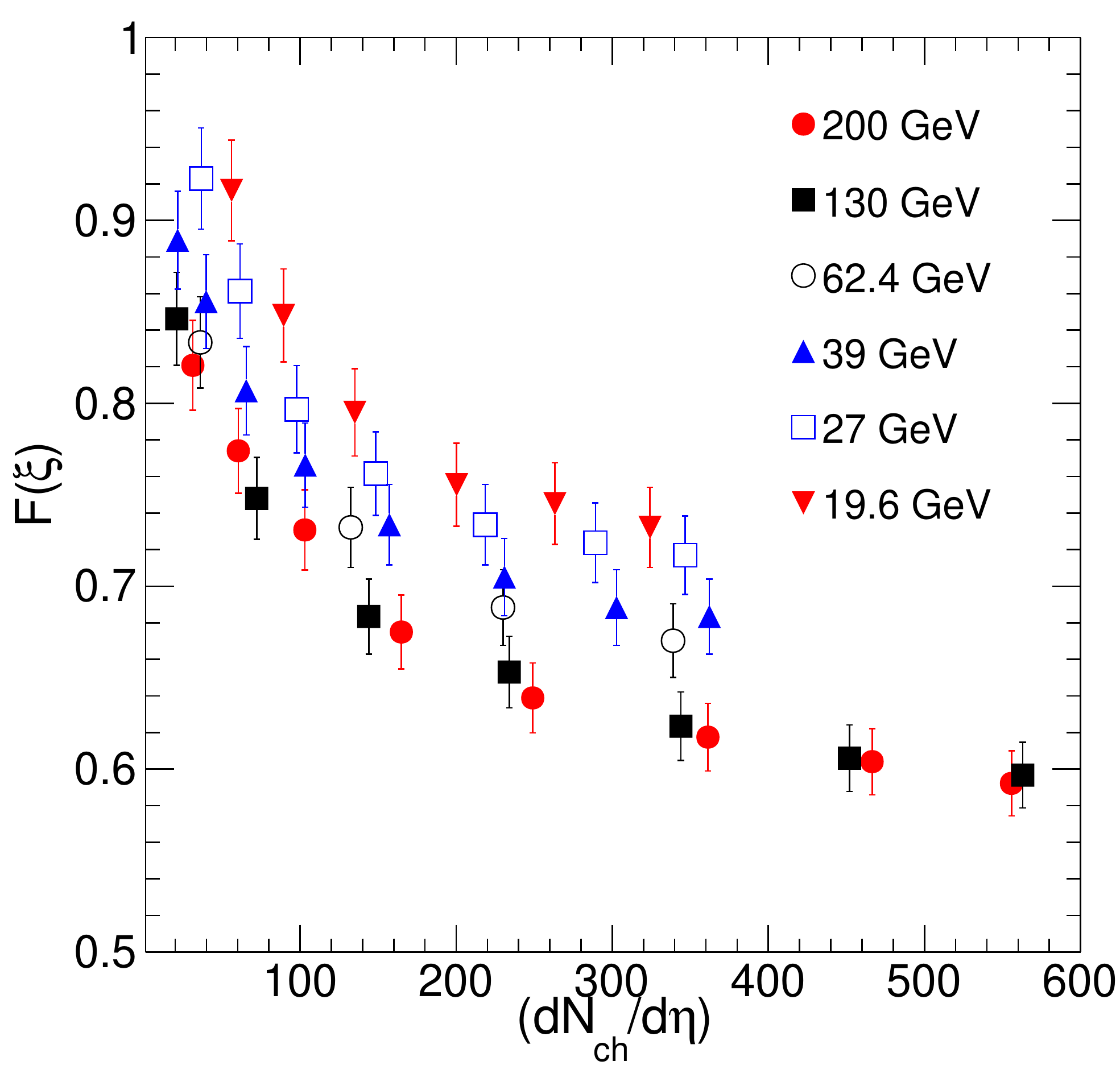}
\caption[]{(Color online) Color Suppression factor $F(\xi)$, as a function of charged particle pseudorapidity density, $dN_{\rm ch}/d\eta$.Different symbols represent different centre-of-mass energies.}
\label{dndetapsi}
\end{figure}

\begin{figure}
\includegraphics[height=20em]{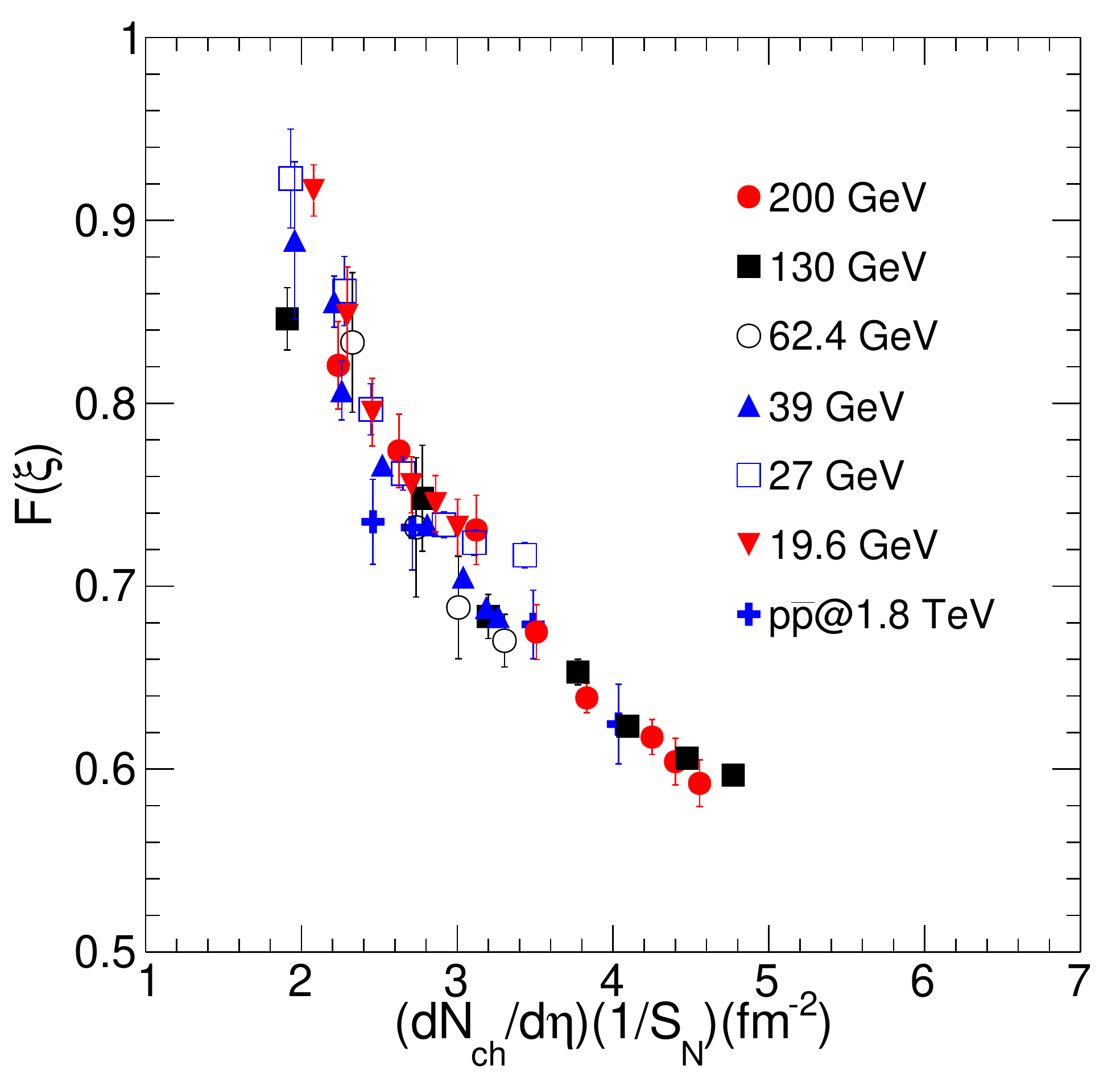}
\caption[]{(Color online) Color Suppression factor $F(\xi)$, as a function of pseudorapidity density divided by the nuclear overlap function, $(dN_{\rm ch}/d\eta)/S_{\rm N}$. Different symbols represent different centre-of-mass energies along with high-multiplicity non-jet $p\bar{p}$ collisions at $\sqrt{s}$= 1.8 TeV at Tevatron~\cite{Gutay:2015cba}.}
\label{dndetapsiFit}
\end{figure}

In figure~\ref{dndetapsi}, we show the extracted values of $F(\xi)$ as a function of pseudorapidity density of the charged particles ($dN_{\rm ch}/d\eta$) for Au + Au collisions at RHIC energies from 19.6 to 200 GeV. $dN_{\rm ch}/d\eta$ values for different centrality classes are taken from experimental data~\cite{star1, star2}. $F(\xi)$ decreases with increasing collision energies as well as centralities. The decreasing trend of $F(\xi)$ is due to the production of high string density, which reveals more suppression of color charges in comparison to lower collisions energies and centralities. $F(\xi)$ in heavy-ion collisions is presented as a function of $dN_{\rm ch}/d\eta$ scaled with the transverse overlap area $S_{\rm N}$, which are estimated by using the Glauber model~\cite{star1} along with the results in high-multiplicity non-jet p$\rm \bar{p}$ collisions at $\sqrt{s}$ = 1.8 TeV from FNAL (Fermi National Accelerator Laboratory) E735 experiment~\cite{Gutay:2015cba} in figure~\ref{dndetapsiFit}. It is observed that $F(\xi)$ falls onto a universal scaling curve for hadron-hadron and nucleus-nucleus collisions. Particularly, in the most central Au+Au collisions and high multiplicity p$\rm \bar{p}$ collisions, $F(\xi)$ values fall in a line. This suggests that the percolation string densities are independent of collision energies and collision systems.

\begin{figure}
\includegraphics[height=20em]{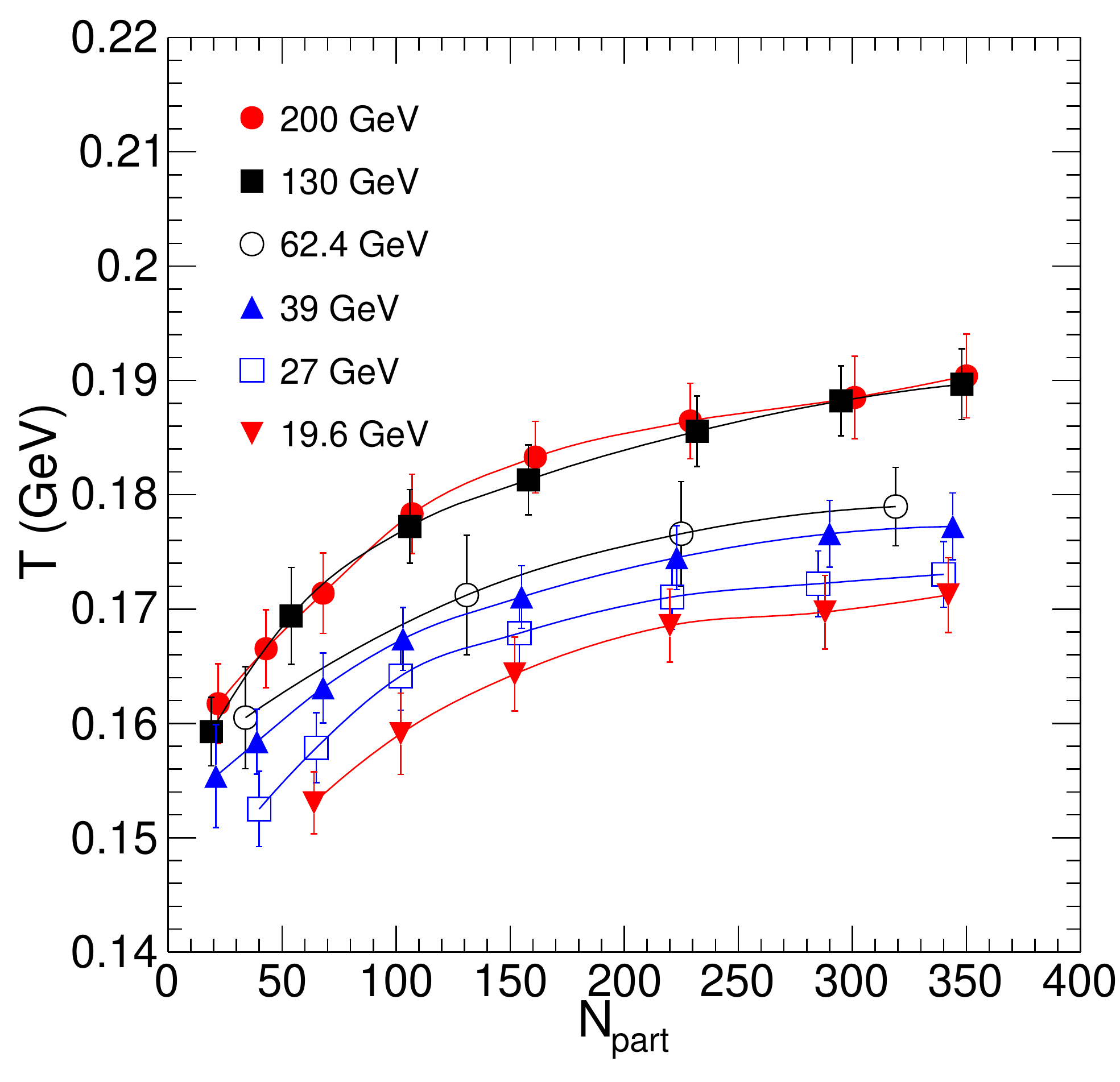}
\caption[]{(Color online) The initial temperatures at RHIC energies estimated in CSPM as a function of number of participants ($N_{part}$) for Au+Au collisions. }
\label{tempNp}
\end{figure}

\begin{figure}
\includegraphics[height=20em]{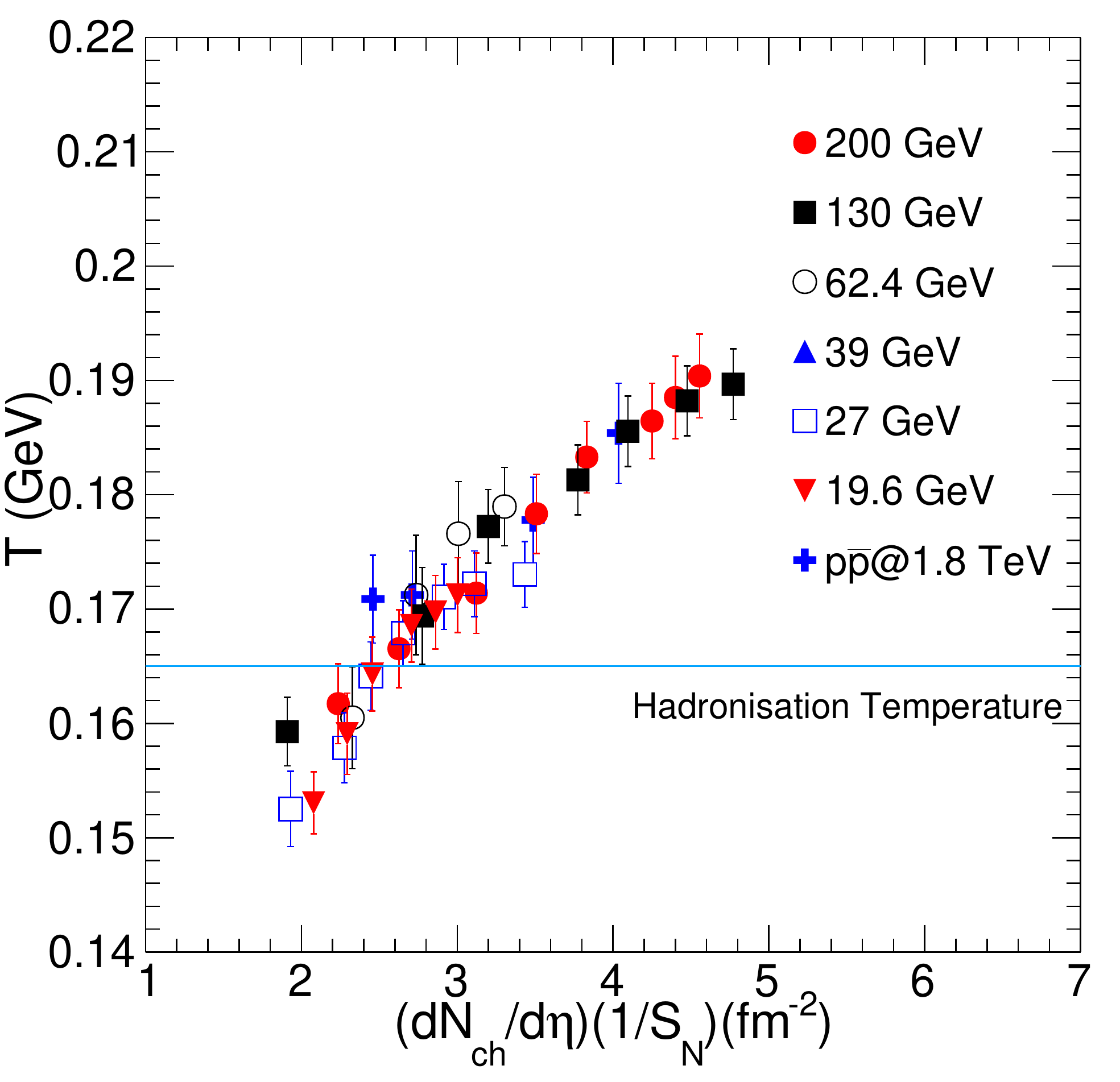}
\caption[]{(Color online) Temperature as a function of $(dN_{\rm ch}/d\eta)/S_{\rm N}$ from pp and Au + Au collisions. The horizontal line corresponds to T $\sim$ 165 MeV, which is the universal hadronisation temperature~\cite{Becattini:2010sk}.}
\label{Tscaled}
\end{figure}

The initial temperature, $T(\xi)$ for different centrality classes is obtained from $F(\xi)$ by using eq.~\ref{eq8} for all the energies. Figure~\ref{tempNp} shows $T(\xi)$ as a function of $N_{part}$. It is found that $T(\xi)$ increases with $N_{part}$ as well as with the centre-of-mass energy. Figure~\ref{Tscaled} represents the initial temperatures versus $dN_{\rm ch}/d\eta/S_{\rm N}$ for various collision energies and systems and we notice that it follows a universal curve for hadron-hadron and nucleus-nucleus collisions. The horizontal line at T $\sim$ 165 MeV is obtained by comparing the hadron yields measured in different collision systems such as pp, A+A, and $\rm e^++e^-$ with the statistical hadron gas model~\cite{Becattini:2010sk}. This temperature corresponds to the critical percolation density, $\xi_c$, where percolation phase transition takes place. The initial temperatures obtained in most and mid-central collisions at RHIC energies and in high multiplicity hadron-hadron collisions at $\sqrt{s}$ = 1.8 TeV are above the hadronisation temperature, which evokes the creation of deconfined matter in these collision energies and systems. 

\begin{figure}
\includegraphics[height=20em]{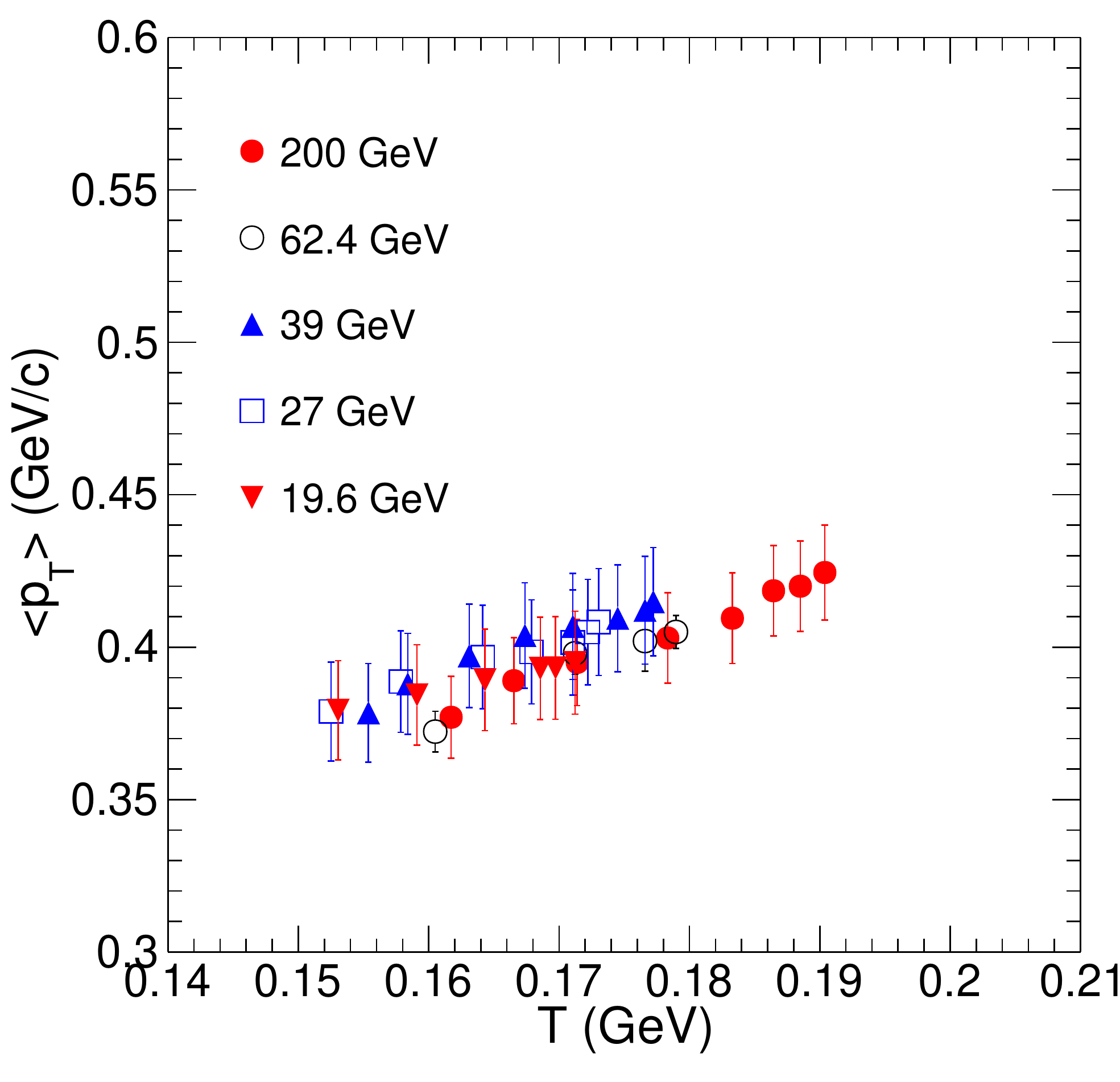}
\caption[]{(Color online) Average momentum $\langle p_{T} \rangle$ of pions as a function of initial temperature, T. $\langle p_{T} \rangle$ are obtained from experimental data~\cite{star1} and initial temperatures are estimated by using CSPM. Different symbols are for different centre-of-mass energies.}
\label{meanpT}
\end{figure}

The mean transverse momentum $\langle p_{T} \rangle$ of pions as a function of initial temperature $T(\xi)$ is shown in figure~\ref{meanpT}. The results are presented for different centralities at RHIC energies. The available $\langle p_{T} \rangle$ values are taken from the experimental data ~\cite{star1, star2} and $T(\xi)$ are calculated using the CSPM as mentioned above. $\langle p_{T} \rangle$ values show a scaling behaviour as a function of the initial temperature for all the colliding energies. We notice that, $\langle p_{T} \rangle$ in the most central collisions of lower energies overlaps with that obtained in the most peripheral collisions of higher energies, which may hint the formation of similar systems in the peripheral collisions at higher energies as formed in the most central collisions at lower energies. Again, this supports the hypothesis of effective energy for particle production~\cite{Sarkisyan,Mishra:2014dta,Sarkisyan-Grinbaum:2018yld,Sarkisyan:2015gca}.

The initial energy density can also be calculated using CSPM for different centrality classes at RHIC energies. Assuming the boost invariant Bjorken hydrodynamics~\cite{Bjorken:1982qr} with CSPM, the initial energy density is expressed as follows,
 
\begin{eqnarray}
 \varepsilon = \frac{3}{2}\frac{\frac{dN_{\rm ch}}{dy}\langle m_{T} \rangle}{S_{\rm N}\tau_{\rm pro}},
 \label{p1}
\end{eqnarray}

where $S_{\rm N}$ is the nuclear overlap area and $\tau_{\rm pro}$, the production time for a boson (gluon), is described by~\cite{taopro,Sahoo:2017umy}

\begin{eqnarray}
\tau_{\rm pro} = \frac{2.405\hbar}{\langle m_{T} \rangle}.
 \label{tau}
\end{eqnarray}

Here, $m_T$ = $\sqrt{m^2+p_T^2}$ is the transverse mass. For evaluating $\varepsilon$, we use the charged pion multiplicity $dN_{\rm ch}/dy$ at mid-rapidity and $S_N$ values for each centrality bin for Au+Au collisions at $\sqrt{s_{\rm NN}}$ = 19.6 - 200 GeV~\cite{star1, star2}.

\begin{figure}
\includegraphics[height=20em]{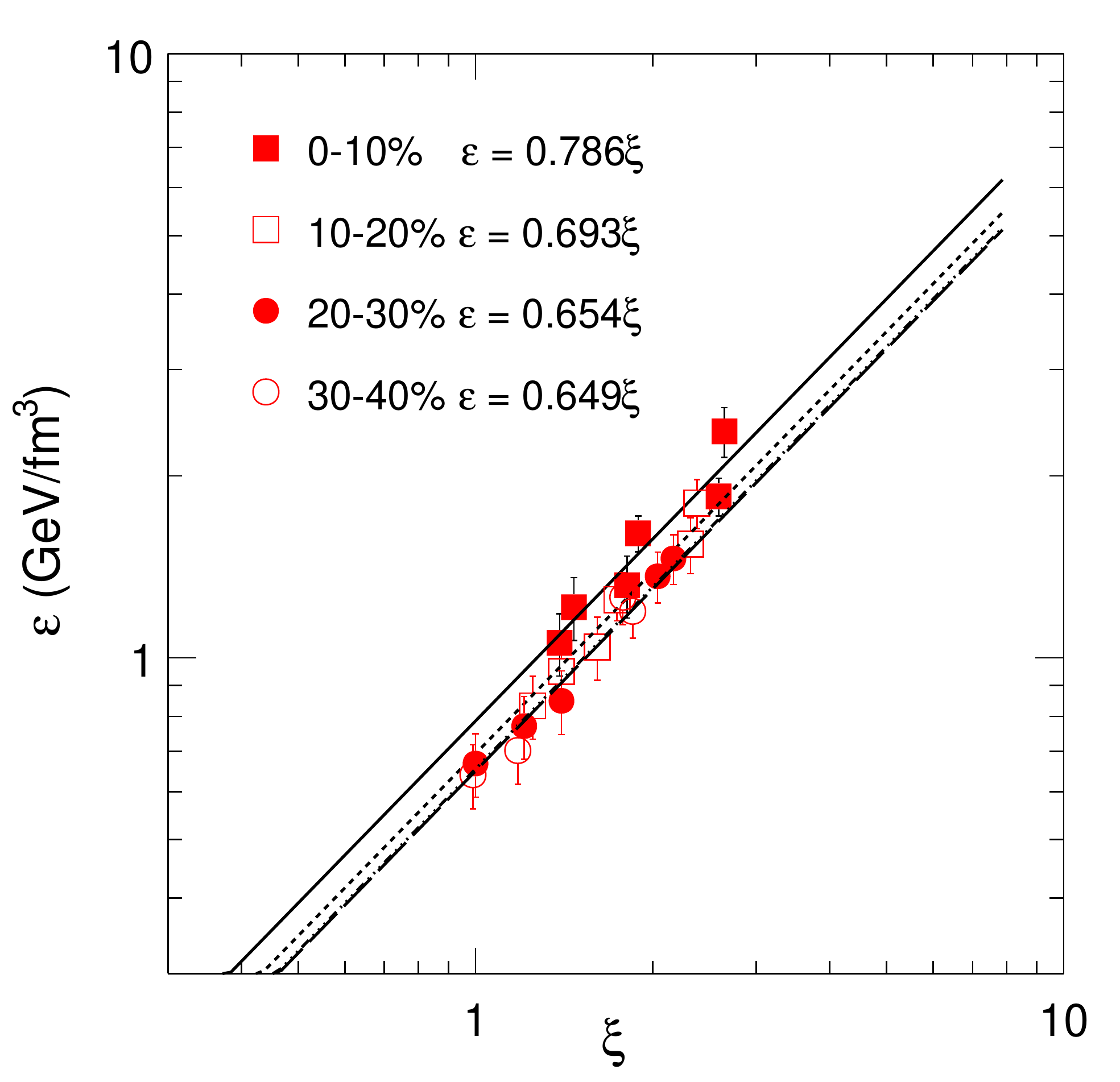}
\caption[]{(Color online) 
Initial energy density, $\varepsilon$ as a function of the percolation density parameter ($\xi$) for different centrality classes at RHIC energies from 19.6 to 200 GeV. The different lines are the linear fits to data for different centralities.}
\label{epsilon}
\end{figure}

Figure~\ref{epsilon} shows the variation of the energy density, estimated by using eq.~\ref{p1} with $\xi$ for different centrality classes at RHIC energies. It is found that $\varepsilon$ is proportional to $\xi$ for all the centrality bins. In our previous work~\cite{Sahoo:2017umy}, we observed that $\varepsilon$ is proportional to $\xi$. Similarly, here we fit the results with a linear function, $\varepsilon$ = A$\times\xi$ (GeV/$fm^3$) for different centrality classes and extract the parameters. The parameter, $A$ = 0.786, 0.693, 0.654, 0.649 (GeV/$fm^3$) for 0-10\%, 10-20\%, 20-30\%, 30-40\% centrality classes, respectively. It shows a decreasing trend from central to peripheral collisions as expected.

\begin{figure}
\includegraphics[height=20em]{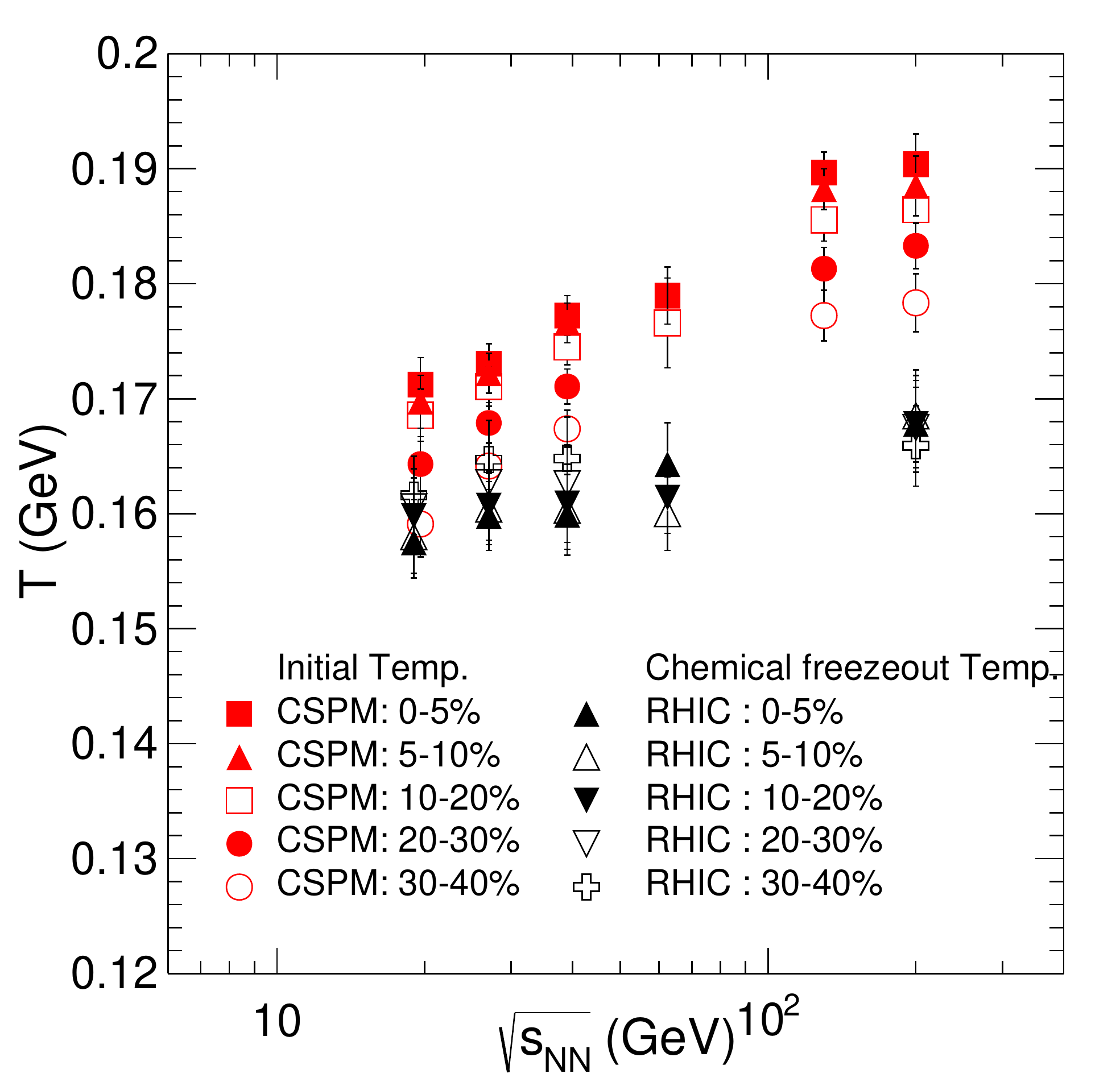}
\caption[]{(Color online) Initial temperature, T estimated by using CSPM at RHIC energies (red symbols) and extracted chemical freeze-out temperatures for Grand Canonical Ensemble using the experimental data at RHIC energies~\cite{star2}  (black symbols) for different centrality classes.}
\label{TSnn}
\end{figure}

Figure~\ref{TSnn} shows the initial temperature estimated by using CSPM at RHIC energies and the  chemical freeze-out temperatures as a function of centre-of-mass energy. The chemical freeze-out temperatures are obtained by fitting the experimental yield of hadrons~\cite{star2} for different centralities in Au+Au collisions at $\sqrt{s_{\rm NN}}$ = 19.6 - 200 GeV. We observe that the initial temperature increases with the collision energies for all the centralities while the chemical freeze-out temperatures remain constant within the uncertainties. Also, the initial temperatures are found close to the freeze-out temperatures at lower energies and the difference between them increases as we move towards the higher collision energies. A strong centrality dependence is found at higher energies in particular, at $\sqrt{s_{\rm NN}}$ = 130 and 200 GeV compared to the lower energies. 

\begin{figure}
\includegraphics[height=20em]{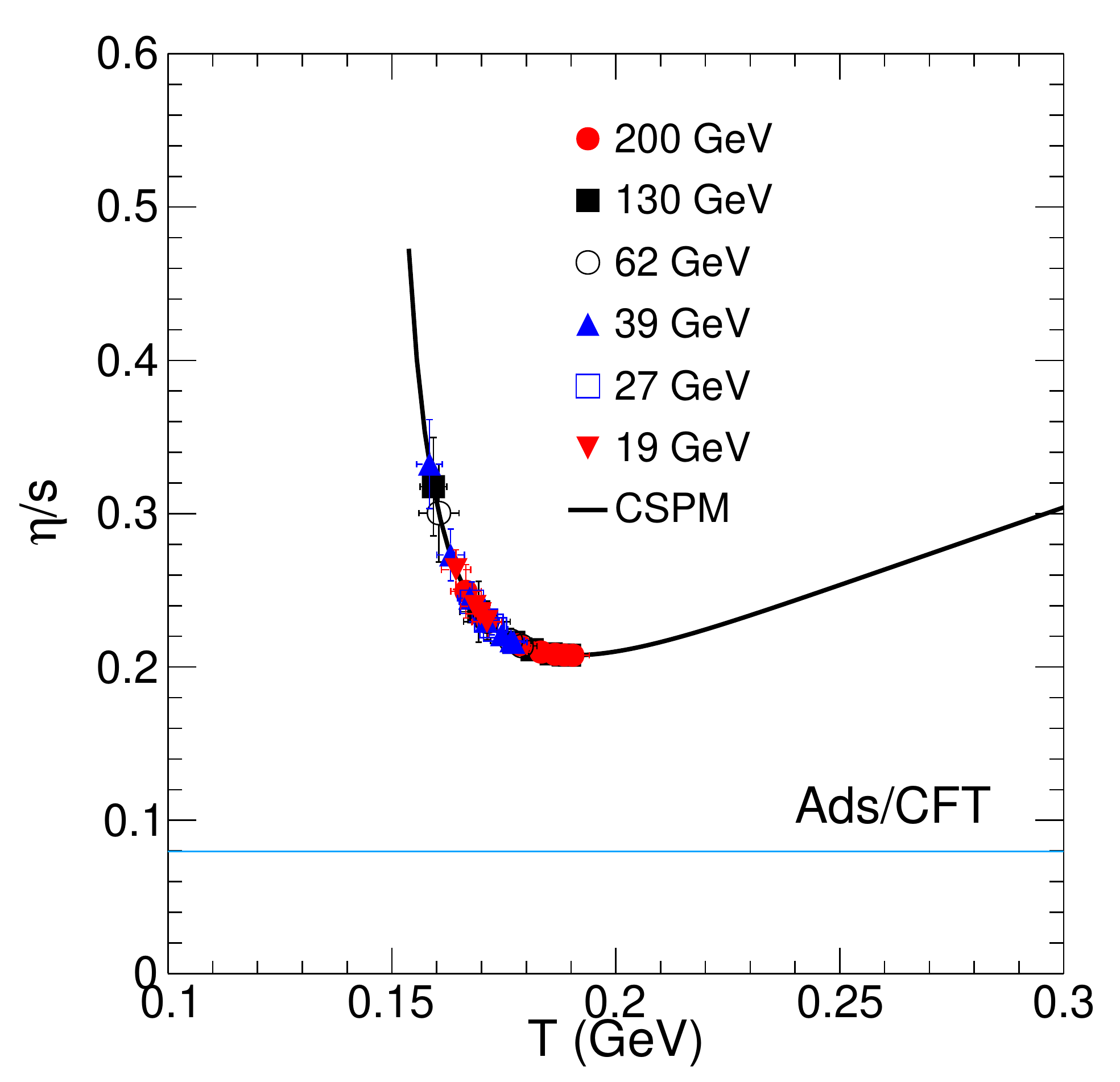}
\caption[]{(Color online) $\eta/s$ as a function of temperature. The markers show the results from various RHIC energies for different centrality bins from the CSPM. The black solid line shows the extrapolation to higher temperatures from CSPM. The solid horizontal line around  $1/4\pi$ represents the AdS/CFT limit~\cite{Kovtun:2004de}.}
\label{EtabyS}
\end{figure}

We first investigate the centrality dependence of shear viscosity to entropy density ratio ($\eta/s$) for various RHIC energies, which act as a measure of the fluidity of the matter formed in heavy-ion collisions. In the framework of the color string percolation model, $\eta/s$ is given by~\cite{Sahoo:2017umy},

\begin{eqnarray}
 \eta/s \simeq \frac{TL}{5(1-e^{-\xi})},
 \label{etas}
\end{eqnarray}
where T is the initial temperature and $L$ is the longitudinal extension of the string $\sim$ 1 fm. We use eq.~\ref{etas} to study the centrality dependence of $\eta/s$ as a function of T as shown in the figure~\ref{EtabyS} for various RHIC energies, where higher T corresponds to the most central collisions and decreases as we go towards peripheral collisions for all energies. The horizontal line shown in this figure is the conjectured lower bound for $\eta/s$ by Ads/CFT calculations~\cite{Kovtun:2004de}. It is perceived that, $\eta/s$ decreases as we go from peripheral to central collisions at all the collision energies and found to be minimum at most central collisions at $\sqrt{s_{\rm NN}}$ = 130 and 200 GeV. 

\begin{figure}
\includegraphics[height=20em]{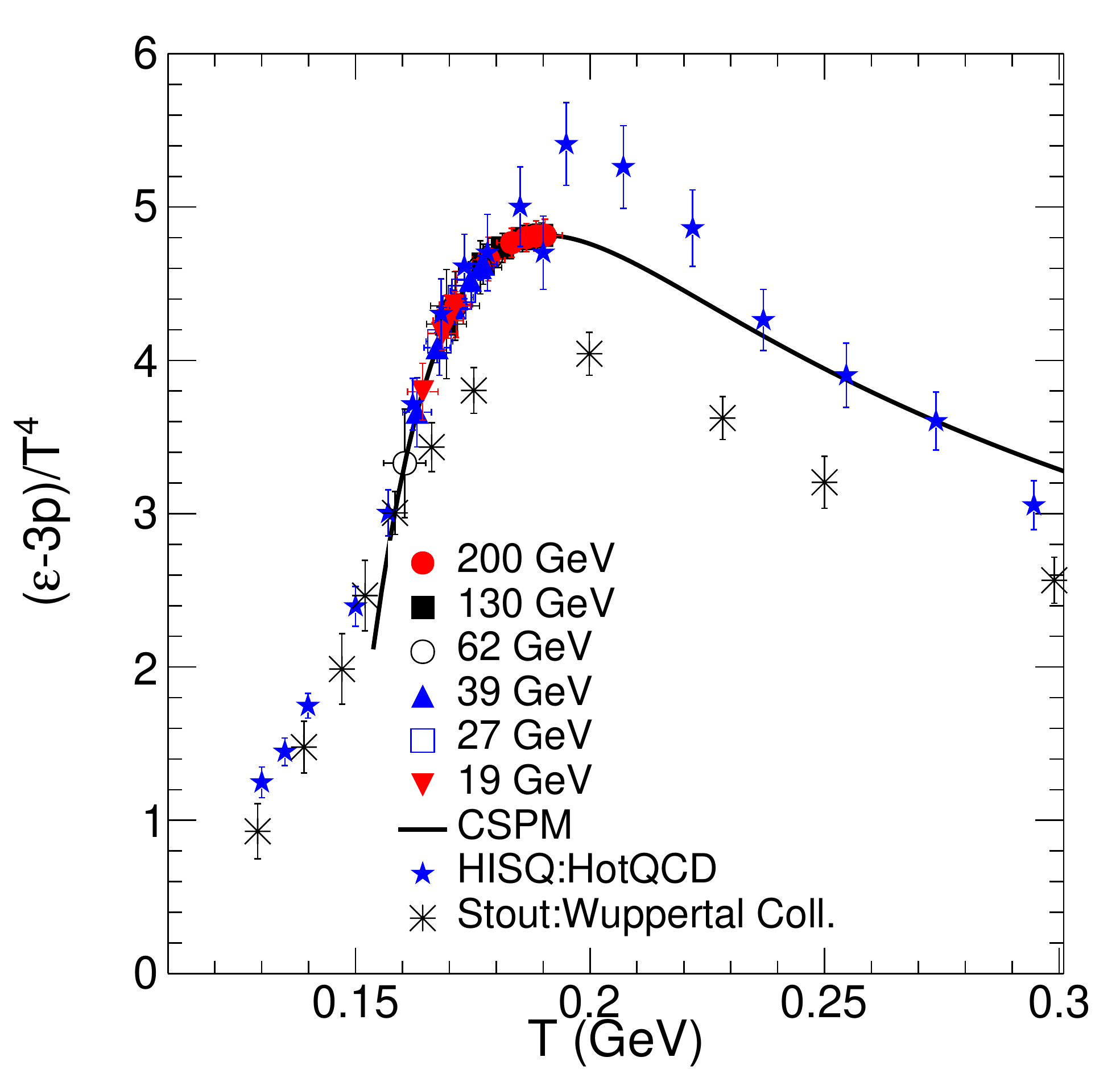}
\caption[]{(Color online) Variation of trace anomaly $\Delta = (\varepsilon-3P)/T^{4}$ with respect to temperature. Blue star markers represent results from HotQCD collaboration~\cite{Bazavov:2014pvz}. Black asterisk  markers refer to Wuppertal collaboration~\cite{Borsanyi:2010cj}. Different markers are the results obtained from CSPM for RHIC energies for different centrality bins and the black line represents the extrapolated CSPM results.}
\label{trace}
\end{figure}

Figure~\ref{trace} shows the centrality dependence of trace anomaly ($\Delta = (\varepsilon-3P)/T^{4}$), which is the reciprocal of $\eta/s$ versus temperature for different RHIC energies. We also compare the CSPM calculations with the LQCD data~\cite{Bazavov:2014pvz,Borsanyi:2010cj} and find that CSPM results are closer to the HOT QCD collaboration results. $\Delta$ increases with centrality for all the RHIC energies and has its maximum value for the most central collisions at $\sqrt{s_{\rm NN}}$ = 130 and 200 GeV.

\section{Summary and Outlook}
\label{summary}
We have presented the calculation of percolation parameters and various thermodynamical observables in nucleus - nucleus collisions at RHIC energies starting from $\sqrt{s_{\rm NN}}$ = 19.6 to 200 GeV as a function of collision centrality. We have extracted $\xi$ and $F(\xi)$ for various centralities by fitting the transverse momentum spectra of charged particles in order to obtain thermodynamical observables. The main findings of this work are summarized as:
\begin{itemize}

\item An occurrence of a phase transition from hadron gas to Quark-Gluon Plasma is described by the critical percolation density in CSPM. We have studied the percolation density parameter as a function of the number of participants at various RHIC energies. We find that, the critical percolation density ($\xi_{c} \sim$ 1.2) is reached for the most central collisions at all the analysed collision energies. However, as we go towards peripheral collisions, the critical percolation density is difficult to achieve at lower collision energies. 

\item We have studied the color suppression factor ($F(\xi)$) in CSPM as a function of pseudorapidity density normalised by nuclear overlap area ($S_{\rm N}$) at RHIC energies. A universal scaling behaviour is observed for all the collision energies. The results from hadron-hadron collisions at $\sqrt{s}$ = 1.8 TeV strengthen the observation showing a similar behaviour as obtained in Au+Au collisions. This indicates that the percolation string densities are independent of collision systems and collision energies.

\item A centrality dependence of initial temperatures is presented for RHIC energies and high multiplicity p$\rm \bar{p}$ collisions at $\sqrt{s}$ = 1.8 TeV. Initial temperatures increase with the number of participants for all the collision energies. We have also shown the initial temperature as a function of $dN_{\rm ch}/d\eta$ scaled by $S_{\rm N}$ and find a universal scaling behaviour both in nucleus-nucleus and hadron-hadron collisions. It is also observed that the initial temperatures obtained in central collisions at RHIC energies and in high multiplicity hadron-hadron collisions at $\sqrt{s}$ = 1.8 TeV are above the hadronisation temperature, which advocates the possibility of the creation of deconfined matter in these collision energies and systems. This also justifies a deep look into the LHC pp high multiplicity events.   

 \item We have presented the average transverse momentum ($\langle p_{T} \rangle$) of pions as a function of temperature for various centralities at RHIC energies. It is found that $\langle p_{T} \rangle$ in the most central collisions of lower energies overlap with that obtained in the most peripheral collisions of higher energies. This suggests that, the system formed in the peripheral collisions at higher energies is similar to that formed in the most central collisions at lower energies, justifying the hypothesis of effective energy.
 
 \item $\eta/s$ as a function of temperature for different centralities are studied in CSPM. A strong centrality dependence of $\eta/s$ is found at RHIC energies. The minimum $\eta/s$ is observed for the most central collisions at $\sqrt{s_{\rm NN}}$ = 130 and 200 GeV, which envisages the formation of perfect fluid at these energies.

\end{itemize}

\section*{ACKNOWLEDGEMENTS}
The authors acknowledge stimulating discussions with Dr. Brijesh K. Srivastava during the preparation of the manuscript. The financial supports from ALICE Project No. SR/MF/PS-01/2014-IITI(G) of Department of Science \& Technology, Government of India is gratefully acknowledged.


\end{document}